\documentclass[%
reprint,
showpacs,
bibnotes,
amsmath,amssymb,
aps,
pra,
floatfix,
]{revtex4-1}

\usepackage{graphicx}
\usepackage{dcolumn}
\usepackage{bm}
\usepackage{hyperref}
\usepackage{multirow}
\usepackage{array}
\usepackage{booktabs}
\usepackage{ctable}
\usepackage{upgreek}
\usepackage{mathrsfs}
\usepackage{amssymb}
\usepackage{amsbsy}
\usepackage{color}
\usepackage{cancel}
\usepackage{marginnote}
\usepackage{amsfonts}
\newcommand{\bcen}{\begin{center}}
\newcommand{\ecen}{\end{center}}
\newcommand{\btab}{\begin{tabular}}
\newcommand{\etab}{\end{tabular}}
\newcommand{\bdes}{\begin{description}}
\newcommand{\edes}{\end{description}}

\newcommand{\beq}{\begin{equation}}
\newcommand{\eeq}{\end{equation}}
\newcommand{\bea}{\begin{eqnarray}}
\newcommand{\eea}{\end{eqnarray}}

\newcommand{\half}{\frac{1}{2}}
\newcommand{\bary}{\begin{array}}
\newcommand{\eary}{\end{array}}
\newcommand{\benum}{\begin{enumerate}}
\newcommand{\eenum}{\end{enumerate}}
\newcommand{\bitem}{\begin{itemize}}
\newcommand{\eitem}{\end{itemize}}

\newcommand{\btau}{\mbox{\boldmath $ \tau $}}
\newcommand{\blam}{{\boldsymbol{\lambda}}}

\newcommand{\bk} { \bm{k} }

\newcommand{\bq} { \bm{q} }
\newcommand{\br} { \boldsymbol{r}}

\newcommand{\bK} { \boldsymbol{K} }

\newcommand{\bzero} { {\boldsymbol{0}}}

\newcommand{\dou}{\partial}

\newcommand{\D}[1]{\mbox{d}{#1}} 
\newcommand{\grad}{\mbox{\boldmath $\nabla$}}

\newcommand{\ket}[1]{| #1 \rangle}
\newcommand{\braket}[2]{\langle #1 | #2 \rangle}

\newcommand{\eqn}[1] {eqn.~(\ref{#1})}
\newcommand{\Eqn}[1] {Eqn.~(\ref{#1})}
\newcommand{\prn}[1] {(\ref{#1})}

\newcommand{\fig}[1]{fig.~\ref{#1}}
\newcommand{\Fig}[1]{Fig.~\ref{#1}}

\makeatletter

\newcommand{\Rmnum}[1]{\expandafter\@slowromancap\romannumeral #1@}
\makeatother

\newlength{\myfigwidth}
\newlength{\myhalffigwidth}
\newcommand{\as}{a_{s}}
\newcommand{\azero}{a_{0}}

\newcommand{\sthre}{{\varepsilon_{th}^s}}

\newcommand{\abg}{{a}_{bg}}
\newcommand{\Binf}{B_\infty}
\newcommand{\Bzero}{B_0}

\newcommand{\eCCB}{\varepsilon_\phi}
\newcommand{\Tm}{\mathbb{T}}
\newcommand{\cG}{{\cal G}}

\newcommand{\mylabel}[1]{\label{#1}} 

\newcommand{\mycite}[1]{\cite{#1}}

\usepackage{lineno}

\begin{document}

\relax



\title{Feshbach Resonance in a Synthetic Non-Abelian Gauge Field}

\author{Vijay B. Shenoy}
\email{shenoy@physics.iisc.ernet.in}

\affiliation{Centre for Condensed Matter Theory, Department of Physics, Indian Institute of Science, Bangalore 560 012, India}



\date{\today}

\begin{abstract}

We study the Feshbach resonance  of  spin-$1/2$ particles in
the presence of a uniform synthetic non-Abelian gauge field
that produces spin orbit coupling along with constant spin
potentials. We develop a renormalizable quantum field theory
that includes the closed channel boson which engenders the Feshbach resonance, in the presence of the gauge field. By a
study of the scattering of two particles in the presence of the gauge field, we show that the
Feshbach magnetic field, where the apparent low energy
scattering length diverges, depends on the conserved centre of
mass momentum of the two particles. For high symmetry gauge
fields, such as the one which produces an isotropic Rashba spin
orbit coupling, we show that the system supports two bound states
over a regime of magnetic fields for a negative background
scattering length and resonance width comparable to the
energy scale of the spin orbit coupling. We discuss the
consequences of these findings for the many body setting, and point out that a broad resonance (width larger than spin orbit coupling energy scale) is most favourable for the realization of the rashbon condensate.

\end{abstract}


\maketitle

Simulation of quantum matter using cold atoms\mycite{Ketterle2008,Bloch2008,Esslinger2010,Cirac2012} has emerged as a very active area of physics research. This owes to the unprecedented tunability that cold atom systems offer in terms of creating hamiltonians with desired one particle levels and interactions. With the recent advances in  synthetic gauge fields\mycite{Lin2009A, Lin2009B, Lin2011,Shanxi2012,MIT2012} the possibility of using cold atoms to simulate even exotic topological states has been significantly enhanced.

These advances have motivated many an effort in the theoretical
study of bosons\mycite{Stanescu2008,Wang2010,Ho2011,Wu2011} and
fermions\mycite{Vyasanakere2011TwoBody,Vyasanakere2011BCSBEC,Yu2011,Hu2011,Gong2011,Iskin2011a,Han2011,Vyasanakere2011Rashbon,Cui2012,Zhang2012,Goldman2012}
in synthetic gauge fields. Uniform non-Abelian SU(2) gauge
fields induce a Rashba-like spin orbit coupling. For fermions
interacting with a contact singlet attraction described by an
energy independent scattering length $\as$, a non-Abelian gauge
field ``amplifies the attractive
interaction''\mycite{Vyasanakere2011TwoBody} rendering the
critical scattering length required for bound state formation
negative. Indeed high symmetry non-Abelian gauge fields
induce a bound state\mycite{Vyasanakere2011TwoBody} for any
scattering length(see also, \mycite{Cui2012,Zhang2012}). For a finite density of fermions, increasing the strength of the spin-orbit coupling $(\lambda)$ induces a crossover from a BCS state comprising of large pairs to a BEC of a new kind of boson, the rashbon, even when the scattering length $\as$ is small and negative. The rashbon is the two fermion bound state obtained with infinite scattering length, and is realized for large $\lambda$ even when $\as$ is small and negative. In other words, the requirement for the realization of the rashbon is that  $|\lambda \as|$ is large. Since $\lambda$ is determined by the lasers used to produce the spin-orbit coupling\mycite{Dalibard2011}, one may use a Feshbach resonance to tune the scattering length be in this desired regime. These considerations, inter alia, provide the natural motivation for this study.

A Feshbach resonance\mycite{Timmermans1999,Chin2010} is
obtained when the bound state of the closed (``triplet'')
channel whose energy is determined by the magnetic field $B$,
crosses the scattering threshold of the open (``singlet'')
channel. The two channels are coupled by the hyperfine
interaction and this produces enhanced scattering of the two particles in
the open channel resulting in a magnetic field dependent
scattering length\mycite{Moerdijk1995}
\beq\mylabel{eqn:aofB}
\azero(B) = \abg \left( 1 - \frac{W}{B- \Binf} \right),
\eeq
where $\abg$ is the background scattering length of the open
channel, $\Binf$ is the field (Feshbach field) which produces a
resonant scattering length, and $W$ is the width of the
resonance. While $\azero(B)$ pertains to particles at the
scattering threshold, scattering at finite energies can be
strongly energy dependent for narrow
resonances\mycite{Chin2010}, and can have interesting effects
in  many body systems.\mycite{Kokkelmans2002,HoCui2012}

In this paper, we develop a renormalizable quantum field theory
of the Feshbach resonance in the presence of a
synthetic gauge field. We obtain explicit expressions for the
fermion scattering $T$-matrix for a generic gauge field. An
important point uncovered by this analysis is that the gauge
field renders the Feshbach field $\Binf$ momentum dependent,
i.~e., $\Binf$ depends on the centre of mass momentum of the
two particles. Further, for high symmetry gauge fields, we show
that in a regime of magnetic field, there are {\em two}
accessible bound states when the background scattering
length is negative and the width is comparable to the spin orbit coupling scale. We discuss the consequences of these results
in the many body setting and the conditions that will enable
the experimental realization of the rashbon condensate.

\medskip
\noindent
{\sc Quantum Field Theory:} Consider a quantum field $\Psi_n(x)$ ($x = (\tau, \br)$, $\tau$
is imaginary time, $\br$ is position vector in 3D) where the
subscript $n=1,...,M (M \ge 3)$ denotes the hyperfine label of
the atom (fermions in this paper). Of particular interest are
the hyperfine labels $n=1,2$, which serve as the two spin
species of interest. The $1$-$2$ interaction in the open (singlet) channel is described by a contact potential $\upsilon$. The
closed (triplet) channel has a bound state described by a
bosonic field $\phi(x)$ called the closed channel boson
(CCB). The CCB, whose energy is $\eCCB(B) = B + B_a$ ($B$ is the
magnetic field, and $B_a$ is an ``adjustment field'', see
below), couples to the singlet density $S(x)$ of the open
channel via hyperfine coupling $\upkappa$. The position
dependent laser coupling of the hyperfine states that produces
the gauge field is denoted by
$H^L_{nn'}(\br)$.\footnote{Generically, $H^L_{nn'}$ will also
  contain the magnetic field $B$. While treating this  is straightforward if
  cumbersome, we keep $H^L$ independent of
  the magnetic field for conceptual clarity.}  At an inverse temperature $\upbeta$, this scenario
is described by the action ($\int \D{x} = \int_0^\upbeta \D{\tau}
\int_\Omega \D{\br}$, $\Omega$ is the volume, repeated indices
summed)
\beq\mylabel{eqn:WithLaser}
\begin{split}
& {\cal S}[\Psi,\phi]  =   \int \D{x} \Psi^\star_n(x) \left( \delta_{n,n'} \left( \frac{\dou}{\dou \tau} - \frac{\grad^2}{2}\right) + H_{nn'}^L(\br) \right) \Psi_n(x) \\
& + \frac{\upsilon}{2} \int \D{x} S^*(x) S(x) + \frac{\upkappa}{\sqrt{2}} \int \D{x} \left(\phi^*(x) S(x) + S^*(x) \phi(x)\right) \\
& + \int \D{x} \phi^*(x) \left( \frac{\dou}{\dou \tau} - \frac{\grad^2}{4} + \eCCB(B) \right) \phi(x).
\end{split} 
\eeq
In the absence of the laser field this reduces to the well
known two channel
model.\cite{Timmermans2001,Duine2004,Braaten2008} Progress is
made by ``locally diagonalizing'' $H_{nn'}^L(\br)$, and
integrating out all but the lowest two states which are mostly
of 1-2 character (see, for example, \mycite{Liu2009}). These two new states are represented by  the quantum fields
$\psi_\sigma(x)$ where $\sigma = \uparrow, \downarrow$. The old 1-2 fields are related to the new ones via $\Psi_n(x) = U_{n\sigma}(\br) \psi_{\sigma}(x)$, where  $U_{n\sigma}(\br), n=1,2$
is a position dependent SU(2) matrix. Writing the action in
terms of the new ``spin-$\half$'' fields results in
\beq\mylabel{eqn:QFT}
\begin{split}
& {\cal S}[\psi,\phi]  =   \int \D{x} \psi^\star_\sigma(x) \left( \delta_{\sigma,\sigma'} \frac{\dou}{\dou \tau} + H_{\sigma \sigma'}(-i \grad, \blam) \right) \psi_\sigma(x) \\
& + \frac{\upsilon}{2} \int \D{x} S^*(x) S(x) + \frac{\upkappa}{\sqrt{2}} \int \D{x} \left(\phi^*(x) S(x) + S^*(x) \phi(x)\right) \\
& + \int \D{x} \phi^*(x) \left( \frac{\dou}{\dou \tau} - \frac{\grad^2}{4} + \eCCB(B) + \varepsilon_{sft} \right) \phi(x).
\end{split}
\eeq
Two points are to be noted. First, the $S(x)$ written in terms
of $\sigma$ states is unchanged from the original. Second, we
have considered the case where the CCB is a deep bound state of
the closed channel potential and hence its wavefunction and
kinetic energy are unaffected by the laser potential apart from a
shift $\varepsilon_{sft}$. Taken together, this leaves
$\upkappa$ unchanged. The term $H_{\sigma \sigma'}(-i
\grad,\blam)$ acting in the open channel now contains the
uniform gauge filed (connection induced by $U_{n\sigma}(\br)$, \mycite{Moody1989}) and any other spin potential
such as the detuning and Zeeman fields\mycite{Liu2009}, all of
which are collectively described by $\blam$. In the following
we set $\varepsilon_{sft}=0$. \Eqn{eqn:QFT} is the quantum
field theory that describes the Feshbach resonance in the
presence of the gauge field. Note that $\upsilon,\upkappa$ and
$B_a$ are the bare coupling constants; the theory requires
renormalization. This is accomplished by considering the two
body problem.

\smallskip
\noindent
{\sc Two body problem:} The one particle eigenstates of $H_{\sigma \sigma'}(-i\grad, \blam)$ are generalized helicity states $\ket{\bk,\alpha} =  \ket{\bk}\otimes\ket{\chi_\alpha(\bk)}$  with dispersion $\varepsilon_\alpha(\bk)$. Here $\ket{\bk}$ is the momentum eigenstate, and $\ket{\chi_\alpha(\bk)}$ is a momentum dependent spin state determined by $\blam$ where  $\alpha = \pm 1$ is the generalized helicity.  Since the action (\eqn{eqn:QFT}) conserves momentum, we can construct two particle states of total momentum $\bq$. The two particle state $\ket{\bq,\bk,\alpha\beta} = \ket{\frac{\bq}{2} + \bk, \alpha} \otimes \ket{\frac{\bq}{2} - \bk, \beta}$ has energy $\varepsilon_{\alpha \beta}(\bq,\bk)=\varepsilon_\alpha(\frac{\bq}{2} + \bk ) + \varepsilon_\beta(\frac{\bq}{2} - \bk )$. Also useful to define is the singlet state $\ket{\bq,\bk,s}$ where the spin structure of the two particles is a singlet. Along with these is the associated   singlet amplitude $A_{\alpha \beta}(\bq,\bk)= \braket{\bq,\bk,s}{\bq,\bk,\alpha \beta}$. The singlet density of states is then obtained as 
\beq\mylabel{eqn:SingletDos}
g_s(\bq,\varepsilon) = \frac{1}{\Omega}\sum_{\bk,\alpha,\beta} |A_{\alpha\beta}(\bq,\bk)|^2 \delta(\varepsilon - \varepsilon_{\alpha \beta}(\bq,\bk)).
\eeq 
A key quantity is the singlet threshold $\sthre(\bq,\blam)$ which is the smallest $\varepsilon$ such that $g_s(\bq,\varepsilon) = 0^+$. In the absence of the gauge field $\sthre(\bq) = \frac{q^2}{4}$. However, in the presence of the gauge field, this is no longer true\mycite{Vyasanakere2011Rashbon}; we will see a specific example  below.

To study the two particle scattering in the open-channel, we solve for the $T$-matrix. Particles with $\ket{\bq,\bk,\alpha\beta} \equiv \ket{\bK}$ are scattered to $\ket{\bq,\bk',\alpha'\beta'} \equiv \ket{\bK'}$. The $T$-matrix element for this process can be calculated as
\beq\mylabel{eqn:RawT}
\Omega\, T_{\bK,\bK'}(\bq,z) = 2 A(\bK) A^*(\bK') \Tm(\bq,z),
\eeq
where $A(\bK) \equiv A_{\alpha \beta}(\bq,\bk)$ and
\beq\mylabel{eqn:TmUnren}
\Tm(\bq,z) = \frac{1}{ (\upsilon + \upkappa^2 G_\phi(\bq,z + \sthre(\bq,\blam))^{-1} - H(z)},
\eeq
where $
G_\phi(\bq,z) = \frac{1}{z - (\frac{q^2}{4} + \varepsilon_\phi(B))} $,
and
\beq\mylabel{eqn:HUnren}
H(z) = \frac{1}{\Omega} \sum_{\bk, \alpha,\beta} \frac{|A_{\alpha \beta}(\bq,\bk)|^2}{z - \varepsilon_{\alpha \beta}(\bq,\bk)}.
\eeq
In eqns.~\prn{eqn:RawT}, \prn{eqn:TmUnren} and \prn{eqn:HUnren}, the energy variable $z$ (in the upper half of the complex frequency space) and $\varepsilon_{\alpha \beta}(\bq,\bk)$ are measured from the singlet threshold $\sthre(\bq,\blam)$. Note that $H(z)$ is a divergent quantity. Regularization, and the concomitant renormalization, is carried out by introducing an ultraviolet momentum cutoff resulting in  $\Lambda = \frac{1}{\Omega} \sum'_{\bk} \frac{1}{k^2}$ (the prime denotes the cutoff). 

To renormalize,  consider the system without the gauge field. Here $H(z) + \Lambda = \frac{\sqrt{-z}}{4\pi}$, the other term in the denominator of \eqn{eqn:TmUnren} can be renormalized as
\beq\mylabel{eqn:arRen}
\left(\upsilon + \frac{\upkappa^2}{z - \eCCB(B)} \right)^{-1} + \Lambda = \frac{1}{4 \pi \abg}\left(\frac{z - (B - \Binf)}{z - (B-\Bzero)} \right)
\eeq
where $B_0 = \Binf + W$. This renormalization is achieved by demanding that the zero energy scattering length (in the absence of the gauge field) reproduces \eqn{eqn:aofB}. In the limit $\Lambda \rightarrow \infty$, the bare parameters are related to the physical parameters via $\frac{1}{\upsilon} + \Lambda = \frac{1}{4 \pi \abg}$, $B_a = B_0$ and $\frac{\upkappa^2}{\upsilon^2} = \frac{W}{4 \pi \abg}$. Note that a necessary condition for renormalizability is that $\abg W > 0$; indeed all the Feshbach resonances tabulated in Table IV of ref.~\cite{Chin2010} satisfy this relation.\footnote{ The necessary condition for the renormalizability is that $s \abg W > 0$ where $s$ is the sign of the moment of the CCB. We have assumed (without loss of generality) that $s=+1$ for our system, i.~e., energy of the CCB increases with increasing $B$}

Armed with the renormalization procedure, we find that
\beq\mylabel{eqn:Tmren}
\Tm(\bq,z) = \frac{1}{\frac{1}{4 \pi a_r(\bq,z)} - \Pi(\bq,z)},
\eeq
where 
\begin{gather}
\frac{1}{4 \pi a_r(\bq,z)}  = \frac{1}{4 \pi \abg}\left(\frac{z - (B - \Binf(\bq,\blam))}{z - (B-\Bzero(\bq,\blam))} \right)\mylabel{eqn:Renar},  \\
\Pi(\bq,z)  = \frac{1}{\Omega} \sum_{\bk} \left[\left( \sum_{\alpha,\beta} \frac{|A_{\alpha \beta}(\bq,\bk)|^2}{z - \varepsilon_{\alpha \beta}(\bq,\bk)} \right) + \frac{1}{k^2}\right], \mylabel{eqn:RenPi}
\end{gather}
with
\beq\mylabel{eqn:Binf}
\Binf(\bq,\blam) = \Binf + \sthre(\bq,\blam) - \frac{q^2}{4},
\eeq
$B_0(\bq,\blam) = \Binf(\bq,\blam) + W$, and $\Binf$ is the Feshbach field in the absence of the gauge field as given in \eqn{eqn:aofB}. \Eqn{eqn:RawT} along with \eqn{eqn:Tmren} provides a complete description of scattering in the open channel across a Feshbach resonance for a generic gauge field.

The result just derived has some very interesting consequences. Note that the nominal Feshbach field $\Binf(\bq,\blam)$ is generally dependent on the centre of mass momentum of the interacting particles! The physics of this owes to the lack of Galilean invariance in the presence of the gauge field\mycite{Zhou2012,Williams2012,Vyasanakere2012Collective}.  Clearly this will have interesting effects in the many body system, particularly at finite temperatures.

We also record here the result for the Green's function for the CCB
\beq\mylabel{eqn:CCBGF}
\cG_\phi(\bq,z) = \frac{1}{z - (B - B_0(\bq,\blam))} \frac{\Tm(\bq,z)}{\Tm_{bg} (\bq,z)},
\eeq
where $\Tm_{bg}(\bq,z)= \frac{1}{\frac{1}{4 \pi \abg} - \Pi(\bq,z)}$. The spectral function of the CCB is obtained as $A_\phi(\bq,\omega) = - \frac{1}{\pi} \Im{\cG_\phi(\bq,\omega^+)}$.

\begin{figure}
\centerline{\includegraphics[width=\myfigwidth]{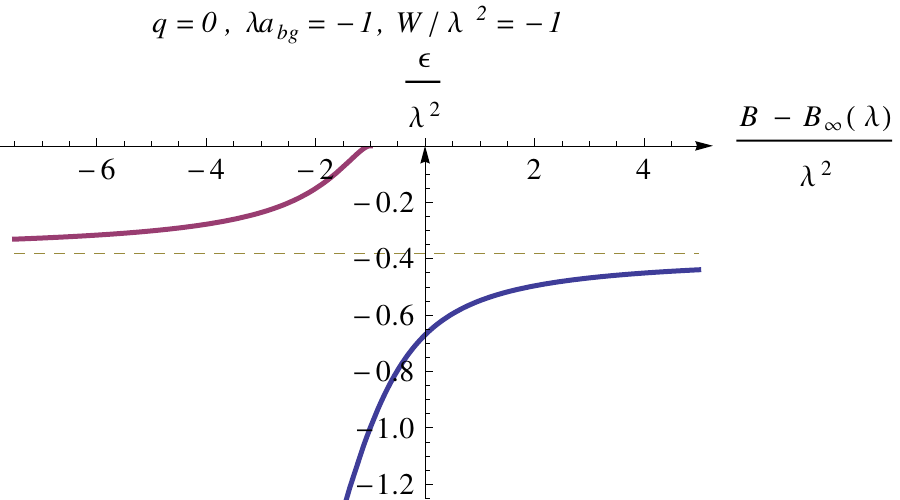}}
\caption{{\bf Bound states across a finite width Feshbach resonance in a spherical gauge field:} For $B \gg \Binf(\lambda)$, there is a single bound state due to the background scattering length. For $B \ll \Bzero(\lambda)$, there is a deep state corresponding to the closed channel boson, and an open channel bound state corresponding, again, to the background scattering length. The magnetic field regime around $B = \Bzero(\lambda)$ is interesting, with {\em two} bound states. The dashed line corresponds to the energy of the bound state $-E_b(\abg)$  (see \eqn{eqn:EneInd}) with an energy independent scattering length $\abg$. }
\mylabel{fig:finW}
\end{figure}

While the results derived above are valid for a generic gauge field, in particular those realized in recent experiments\mycite{Shanxi2012,MIT2012}, we now discuss Feshbach resonance in the spherical gauge field\mycite{Vyasanakere2011TwoBody} where many analytical results are possible.  This gauge field results in an isotropic Rashba spin orbit coupling such that $H_{\sigma \sigma'} = -\delta_{\sigma \sigma'} \frac{\grad^2}{2} + i \lambda \grad \cdot \btau_{\sigma \sigma'}$, where $\btau$ is the vector of Pauli matrices. For an energy independent scattering length $\as$, two particles always have a bound state\mycite{Vyasanakere2011TwoBody} with binding energy 
\beq\mylabel{eqn:EneInd}
E_b(\as) = \frac{1}{4} \left(\frac{1}{\as} + \sqrt{\frac{1}{\as^2} + 4 \lambda^2} \right)^2,
\eeq
and the rashbon binding energy is $\lambda^2$.

The singlet threshold for this gauge field\mycite{Vyasanakere2011Rashbon} has a very interesting character. For $q=|\bq| \le 2 \lambda$, $\sthre(q,\lambda) = - \lambda^2$, i.~e., independent of $q$, and $q > 2 \lambda$, the threshold increases with increasing $q$. Focussing on the regime $q < 2 \lambda$, we find from \eqn{eqn:Binf} that
\beq
\Binf(q,\lambda) = \Binf - \lambda^2 - \frac{q^2}{4},
\eeq
which clearly demonstrates the $q$-dependence of the Feshbach field. In the remainder of the paper, we will discuss the Feshbach resonance at $q=0$ in the spherical gauge field for which  $\Pi(q=0,z)$ has a nice analytic expression
\beq
\Pi(z) = \frac{1}{4 \pi} \left(\sqrt{-z} - \frac{\lambda^2}{\sqrt{-z}} \right).
\eeq

\begin{figure}
\centerline{\includegraphics[width=\myfigwidth]{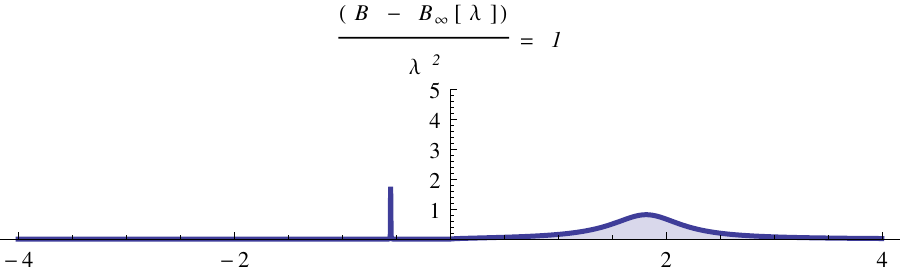}}
\centerline{\includegraphics[width=\myfigwidth]{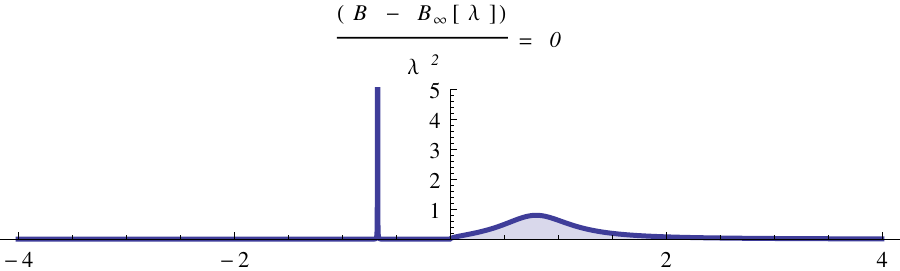}}
\centerline{\includegraphics[width=\myfigwidth]{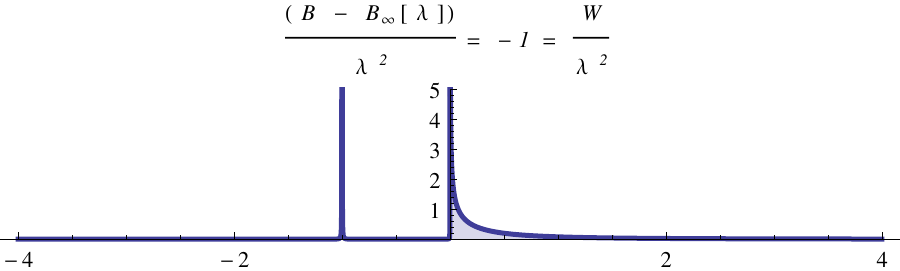}}
\centerline{\includegraphics[width=\myfigwidth]{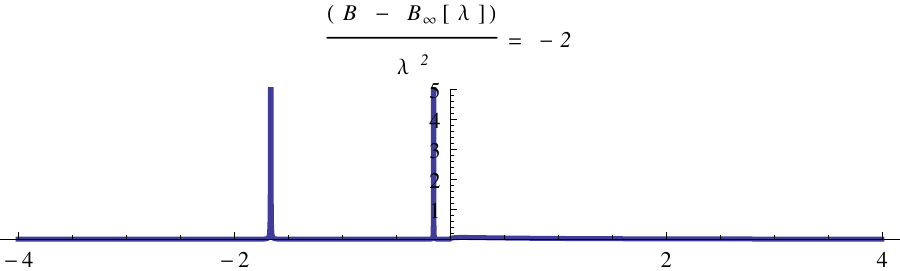}}
\centerline{\includegraphics[width=\myfigwidth]{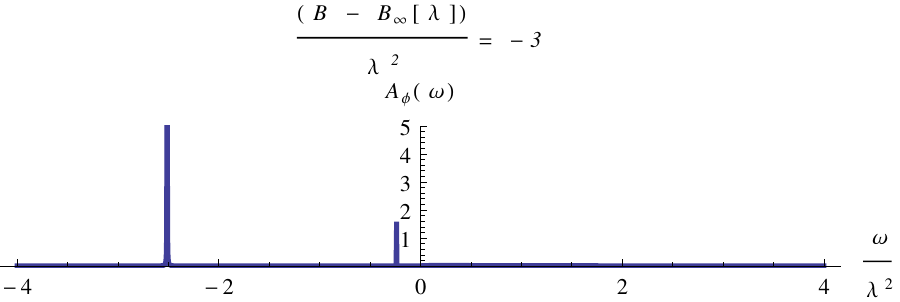}}
\caption{{\bf Evolution of the spectral function of the closed channel boson(CCB):} Physical parameters are same as those in \Fig{fig:finW}. For $B \gg \Binf(\lambda)$, the CCB has most weight in the scattering continuum, while for $B \ll \Bzero(\lambda)$, the CCB does not significantly couple to the open channel. $B \sim \Bzero(\lambda)$ is of interest where the CCB has nearly equal weights in both the bound states. }
\mylabel{fig:spectralCCB}
\end{figure}

\medskip\noindent
{{\sc Finite width resonance:}} We first consider a finite width resonance with a negative
background scattering length. \Fig{fig:finW} shows the energy
of the bound states as the magnetic field is swept across the
resonance. When $B \gg \Binf(\lambda)$ there is a single bound state
which is open channel dominated. Indeed the energy of this
state is slightly below the value given by \eqn{eqn:EneInd}
with $\as = \abg$ (dashed line in \Fig{fig:finW}) owing to the
``level repulsion'' with the closed channel boson. Quite
interestingly, there is just this state at $B=
\Binf(\lambda)$. A second bound state appears only for $B <
\Bzero(\lambda)$, and in the regime of magnetic field around
$\Bzero(\lambda)$ the system supports {\em two} bound
states. The bound state spectrum has the structure of an
avoided crossing between the open channel bound state due to
the background scattering length and the CCB. Further
understanding of the physics can be obtained by a study of
evolution of the spectral function of the CCB shown in
\fig{fig:spectralCCB}. For $B \gg \Binf(\lambda)$, the CCB
resides in the scattering continuum hybridizing with the open
channel states. For $B \ll \Bzero(\lambda)$ the CCB decouples
from the open channel (see lowest panel in
\fig{fig:spectralCCB}). What is noteworthy is  that for $B
\lessapprox \Bzero(\lambda)$, the CCB has nearly equal weight
both the bound states. It will be interesting to explore systems
where these two bound states are accessible, i.~e., close in
energy compared to, e.~g., temperature. 

A system with a positive background scattering length will also produce  qualitatively similar physics. The notable difference is that the bound state induced by the background scattering length is deeper, and the two bound states induced by the resonance will be well separated and likely inaccessible in experiments.

We now turn to the possibility of realization of the rashbon state in a system with a resonance width comparable to the energy scale of the spin-orbit coupling ($\lambda^2$). \Fig{fig:finW} shows that the energy of the bound state at $B=\Binf(\lambda)$ does {\em not} correspond to the rashbon energy of $-\lambda^2$. Furthermore, for $B \lesssim \Binf(\lambda)$, the closed channel character of the bound state increases. It is therefore clear that the rashbon state is not realized across a finite width resonance (of width comparable to the energy scale of the spin orbit coupling).

\begin{figure}
\centerline{\includegraphics[width=\myfigwidth]{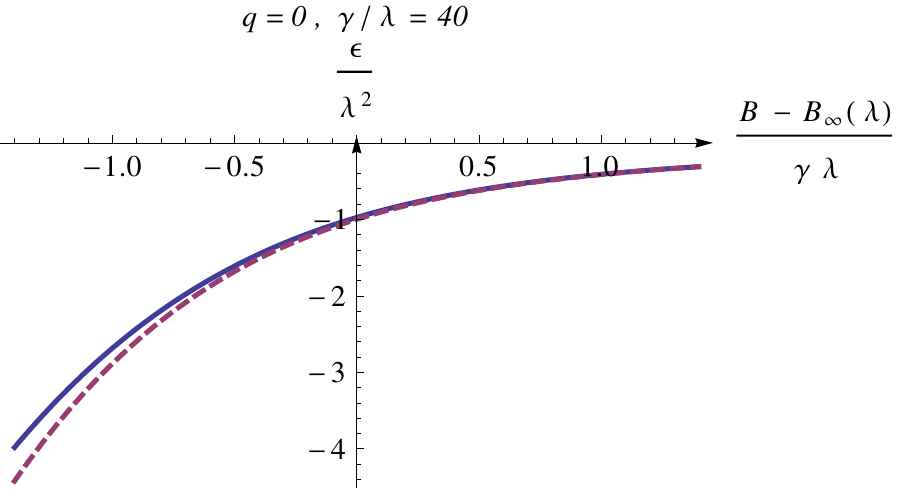}}
\caption{{\bf Bound state across the broad resonance in a spherical gauge field:} Bound state energy of the single bound state (solid line), compared with the result (dashed line) based on an energy independent scattering length (\eqn{eqn:EneInd}). Note that the rashbon state is realized at $B=\Binf(\lambda)$. }
\mylabel{fig:broad}
\end{figure}

\medskip\noindent
{\sc Broad Feshbach Resonance:} A broad resonance is obtained when $\abg \rightarrow 0^-$ and $W \rightarrow -\infty$ keeping $\abg W = \gamma$ finite. For a generic gauge field, in this limit, \eqn{eqn:arRen} becomes
\beq\mylabel{eqn:broad}
\frac{1}{4 \pi a_r(\bq,z)} = \frac{z - (B - \Binf(\bq,\blam))}{4 \pi \gamma}
\eeq
 and its energy dependence can be mitigated when $\gamma$ is large. 

\Fig{fig:broad} shows the bound state spectrum for the spherical gauge field in a broad Feshbach resonance. The key point to be noted is that the system now has only one bound state. In fact, the bound state energy closely matches the energy obtained from \eqn{eqn:EneInd} using $\as$ as the scattering length at zero energy obtained from \eqn{eqn:broad}. What is heartening is that the bound state at $B =\Binf(\lambda)$ does correspond to the rashbon state with weight dominantly in the open channel. Our study clearly points out that a broad resonance, i.~e., whose width is much larger compared to the spin orbit coupling energy scale ($\lambda^2$), is the most favourable system to realize the rashbon.

\medskip\noindent
{\sc Discussion:} The results obtained here have many interesting consequences in the many body setting which we now discuss. Firstly, the $\bq$ dependent shift of the Feshbach field should produce  interesting effects in the many body system. On one hand, the effects of Pauli blocking inhibits bound state formation near $\bq \approx \bzero$ (see \mycite{Shenoy2011UpperBranch}), while the effects of Pauli blocking are minimal for larger $\bq$. On the other hand, the gauge field which promotes bound state formation at small $\bq$ actually inhibits bound state formation\mycite{Vyasanakere2011Rashbon,Shenoy2012FlowSF,PengZhang2012SR} at larger values of $\bq$. The effect of the $\bq$ dependent Feshbach field is therefore not obvious (atleast to the author) -- this is clearly a very interesting problem for further investigation.

Although narrow resonances (width comparable to the spin orbit coupling scale) are not favourable for the realization of the rashbon, they do offer new possibilities. The regime of magnetic fields where there are two bound states is fertile with interesting new physics. In a low density system at low temperatures, the presence of the two bound states will promote fluctuations and possibly inhibit condensation -- a study of this competition is also be an interesting direction for further investigation. The renormalizable field theory developed in this paper could be used for these investigations.

From the point of view of experiments, this work clearly points to the conditions favourable for the realization of the rashbon condensate. What is unequivocally clear from this work and the cited literature is that cold atoms in synthetic gauge fields is a treasure trove of interesting physics. We hope this motivates experimental efforts towards uncovering these.

\noindent
{\bf Acknowledgement:} This work is generously supported  by DAE (SRC grant) and DST, India. The author is grateful to J.~Vyasanakere for discussions, S.~K.~Ghosh and A.~Agarwala for comments on the manuscript.

\bibliography{bibliography_feshbach_sgf}

\end{document}